\documentclass[aps,prl,twocolumn,showpacs,superscriptaddress,groupedaddress,preprintnumbers]{revtex4}  % for review and submission
\usepackage{graphicx}  % needed for figures
\usepackage{dcolumn}   % needed for some tables
\usepackage{bm}        % for math
\usepackage{amssymb}   % for math
\usepackage[latin1]{inputenc}
\usepackage{amsmath}
\usepackage{amsfonts}
\usepackage{amssymb}
\usepackage{setspace}
\usepackage{mathtools}
\usepackage{pstricks}
\usepackage{color}
\usepackage{hyperref}
\usepackage{lipsum}%for mock text
\usepackage{marginnote}% for notes on the marginal

\newcommand{\showarttu}{}
\ifx\showarttu\undefined

\newcommand{\arttupois}[1]{}
\else
\usepackage[normalem]{ulem}
\usepackage{color}

\fi

\providecommand{\f}[2]{\frac{{#1}}{{#2}}}

\newcommand{\ee}[1]{\begin{equation}#1\end{equation}}
\newcommand{\ea}[1]{\begin{align}#1\end{align}}

\begin{document}
 
\preprint{HIP-2015-21/TH, Imperial-TP-2015-TM}
\title{Spacetime curvature and Higgs stability after inflation}

\affiliation{Niels Bohr International Academy and Discovery Center, Niels Bohr Institute,
University of Copenhagen%, Blegdamsvej 17, 2100 Copenhagen, Denmark
}
\affiliation{Department of Physics, Imperial College London%, SW7 2AZ, United Kingdom
}
\affiliation{Department of Physics, King's College London%, Strand, London WC2R 2LS, UK}
}
\affiliation{Helsinki Institute of Physics and Department of Physics, University of Helsinki%, P. O. Box 64, FI-00014, Finland
}
\affiliation{Department of Physics, University of Jyv\"{a}skyl\"{a}
}
\author{M.~Herranen}\email{herranen@nbi.ku.dk}\affiliation{Niels Bohr International Academy and Discovery Center, Niels Bohr Institute,
University of Copenhagen%, Blegdamsvej 17, 2100 Copenhagen, Denmark
}
\author{T.~Markkanen}\email{tommi.markkanen@kcl.ac.uk}\affiliation{Department of Physics, Imperial College London%, SW7 2AZ, United Kingdom
}\affiliation{Department of Physics, King's College London%
}
\author{S.~Nurmi}\email{sami.t.nurmi@jyu.fi}\affiliation{Helsinki Institute of Physics and Department of Physics, University of Helsinki}
\affiliation{Department of Physics, University of Jyv\"{a}skyl\"{a}
}
\author{A.~Rajantie}\email{a.rajantie@imperial.ac.uk}\affiliation{Department of Physics, Imperial College London%, SW7 2AZ, United Kingdom
}\affiliation{Department of Physics, King's College London%
}
\date{\today}

\begin{abstract}

We investigate the dynamics of the Higgs field at the end of inflation in the minimal scenario consisting of an inflaton field coupled to the Standard Model only through the non-minimal gravitational coupling $\xi$ of the Higgs field. Such a coupling is required by renormalisation of the Standard Model in curved space, and in the current scenario also by vacuum stability during high-scale inflation. We find that for $\xi\gtrsim 1$, rapidly changing spacetime curvature at the end of inflation leads to significant production of Higgs particles, potentially triggering a transition to a negative-energy Planck scale vacuum state and causing an immediate collapse of the Universe.
 
\end{abstract}

\pacs{98.80.Cq, 04.62.+v}
\maketitle
The Standard Model (SM) of particle physics can be consistently extrapolated to the Planck scale without any new physics, but the
current measurements of the Higgs boson and top quark masses suggest that the current vacuum state of the Universe would then not be stable.
This instability depends sensitively on the top mass $m_t$, which is subject to significant experimental and theoretical uncertainty \cite{Bednyakov:2015sca}, but
for the best fit values, the Higgs potential turns negative above the instability scale $\Lambda_I\sim 10^{11}$GeV \footnote{This number is a gauge dependent quantity \cite{Jackiw:1974cv}}.
This implies that the current vacuum would eventually decay into a negative-energy Planck scale true vacuum, but its lifetime exceeds the age of the universe by a wide margin \cite{Degrassi:2012ry}.

Whether such a metastable universe could have survived the cosmological evolution, especially inflation, has recently attracted significant interest~\cite{Kobakhidze:2013tn,Herranen:2014cua,Kearney:2015vba}.
In most of the simplest models of inflation, the Hubble rate during inflation is comparable to the current upper bound $H\lesssim 9\times 10^{13}~{\rm GeV}$~\cite{Ade:2015tva}. It may therefore well be above the instability scale, in which case production of Higgs fluctuations could push the field over the potential barrier into the true Planck-scale vacuum~\cite{Kobakhidze:2013tn}. This instability problem is exacerbated by spacetime curvature induced running of the couplings, which makes the Higgs self-coupling negative even at low field values~\cite{Herranen:2014cua,Markkanen:2014poa,Kearney:2015vba}.

Notably, vacuum stability can still be
maintained even during inflation without any new physics coupled to
the SM fields \cite{Herranen:2014cua}, thanks to the Higgs-curvature
coupling $\xi R\hat{H}^\dagger\hat{H}$. This coupling is inevitably generated by radiative
corrections and when assuming the SM to be valid up to the Planck scale it is the only relevant new term when probing sub-Planckian scales. The current experimental constraints are extremely weak, $|\xi|\lesssim 2.6\times 10^{15}$~\cite{Atkins:2012yn}.
With a positive coupling, this term increases the height of the potential barrier between the vacua, thereby increasing the lifetime of the metastable vacuum.
Vacuum stability is maintained for all inflationary scales compatible with the tensor bound
\cite{Ade:2015tva}, provided the electroweak scale value of the running
coupling $\xi(\mu)$ lies above $\xi_{\rm EW}\gtrsim 0.1$
\cite{Herranen:2014cua}.

In this letter, we investigate the instability problem at the end of inflation,
again assuming no new physics or higher-dimensional operators but taking the gravitational coupling $\xi$ into account. In contrast with the Higgs inflation scenario \cite{Bezrukov:2007ep}, we assume that the Higgs is subdominant, and for simplicity we ignore a possible direct coupling to the inflation. We focus on the parametric region $\xi > \xi_c = 1/6$ in which case the Higgs is effectively massive during inflation and does not get displaced from the SM vacuum.
%, unlike in \cite{DeSimone:2012qr}. 
When inflation ends, the curvature scalar $R$ drops rapidly, reducing the height of the potential barrier.
Eventually, a new barrier is generated by the thermalized degrees of freedom re-stabilising the vacuum. During the intermediate period rapid changes in the scalar curvature $R$ can lead to non-adiabatic evolution, giving rise to significant excitations of the Higgs field, and potentially triggering the instability. Two major sources of excitations are a sudden drop in $R$ when inflation ends \cite{Ford:1986sy}, and parametric resonance from the oscillating curvature term sourced by the inflaton~\cite{Bassett:1997az,Tsujikawa:1999jh}.
%. The latter gives very similar effects to standard preheating \cite{Kofman:1994rk} and has been discussed in \cite{Bassett:1997az,Tsujikawa:1999jh}.

We can write the Standard Model Higgs Lagrangian as
\ee{{\cal L}=\vert D_{\mu}\hat{H}\vert^2-\left(-M^2+\xi R\right)\vert \hat{H} \vert^2-\lambda \vert \hat{H} \vert^4\, ,}
where $D_\mu$ is the standard $SU(2)\times U(1)$ covariant derivative generalized to curved space, $M$ the Higgs field mass parameter and $\xi$ the nonminimal coupling. {We will parametrize} the complex Higgs field doublet $\hat{H}$  as 
\ee{
\hat{H}= \f{1}{\sqrt{2}}\left(\begin{array}{c} \hat{h}_1 + i \hat{h}_2 \\ \hat{h}_3 + i \hat{h}_4\end{array}\right)\,. }

Assuming the Friedmann-Robertson-Walker metric $ds^2=dt^2-a^2d\mathbf{x}^2=a^2(d\eta^2-d\mathbf{x}^2)$, and using the conformal time coordinate $\eta$, we write each of the four components $\hat{h}_i$ as
\ee{\hat{h}=\int \f{d^{3}\mathbf{k}}{\sqrt{(2\pi)^3a^2}}\big [\hat{{a}}_\mathbf{k}^{\phantom{\dagger}} \,{f(\eta)}{e^{i\mathbf{k}\cdot \mathbf{x}}}+\hat{{a}}^\dagger_\mathbf{k}  \,{f^*(\eta)}{e^{-i\mathbf{k}\cdot \mathbf{x}}}\big]\, ,}
where $\mathbf{k}$ is the comoving momentum and
$[\hat{a}_{\mathbf{k}}^{\phantom{\dagger}},\hat{a}_{\mathbf{k}'}^\dagger]=\delta^{(3)}(\mathbf{k}-\mathbf{k}'),~~[\hat{a}_{\mathbf{k}}^{\phantom{\dagger}},\hat{a}_{\mathbf{k}'}^{\phantom{\dagger}}]=[\hat{a}_{\mathbf{k}}^{{\dagger}},\hat{a}_{\mathbf{k}'}^\dagger]=0.$  
Ignoring interactions and the Minkowski space mass term $M^2$, we obtain the mode equation
\ee{{f}''(\eta)+ \left[{\mathbf{k}^2}+\left(\xi-\f{1}{6}\right)a^2R \right]f(\eta)=0\, ,\label{eq:mode}}
where the prime denotes $d/d\eta$ and $R=6a''/a^3$. 

We can see that the mode has an effective curvature-induced mass term
\ee{m^2_{\rm curv}=\left(\xi-\f{1}{6}\right)a^2R.}
If the frequency of the mode $\omega^2=\mathbf{k}^2+m_{\rm curv}^2$ satisfies the adiabaticity condition $\left\vert{\omega'{}^2}/{\omega^4}\right\vert \sim\left\vert{{\omega''}}/{\omega^3}\right\vert\ll1$, the mode equation
 may be solved adiabatically via a WKB-type ansatz
\ee{f(\eta)=\f{e^{-i \int^\eta d\eta' \omega}}{\sqrt{2\omega }}.\label{eq:WKB}}
The occupation number of a mode is an adiabatic invariant \cite{Parker:1968mv}, so if a mode starts in the vacuum and evolves adiabatically, it will stay unexcited. Only for a fast change in $R$ do we expect to find sizeable excitations.

For simplicity, we will first assume that the equation of state $w=p/\rho$ changes discontinuously from $w=w_{\rm in}=-1$ to a constant value $w=w_{\rm out}\ge 0$ at the end of inflation. For constant $w$, the scale factor behaves as
%\ee{
$a=\left({\eta}/{\eta_0}\right)^{{2}/{(3w+1)}},
$
%\label{eq:w}}
and the scalar curvature expressed in terms of the Hubble rate $H=\dot{a}/a=a'/a^2$ is $R=3(1-3w)H^2.$
We assume that $\xi\gtrsim 1$, so that the adiabaticity condition is satisfied on both sides of the transition, but at the transition itself the curvature mass term drops instantaneously by a factor
\ee{\frac{m^2_{\rm out}}{m^2_{\rm in}}=\frac{1-3w_{\rm out}}{4}.\label{eq:massratio}}
This non-adiabatic change produces excitations of the Higgs field.

Following \cite{Bernard:1977pq} we can write two sets of solutions: $f_{\rm in}(\eta)$ that reduces to the incoming adiabatic mode in the past and $f_{\rm out}(\eta)$ that reduces to the outgoing mode in the future
with
$\omega_{\rm in}^2={\mathbf{k}^2}+m^2_{\rm in}$ and $\omega_{\rm out}^2={\mathbf{k}^2}+m^2_{\rm out}$, respectively.
The two vacua can be connected with a Bogoliubov transformation  
%\ee{
$f_{\rm in}(\eta)=\alpha_\mathbf{k} f_{\rm out}(\eta)+\beta_\mathbf{k} f^*_{\rm out}(\eta)\,,
$%}
 with Bogoliubov coefficients
\ee{
\alpha_\mathbf{k}=\f{1}{2}\bigg[\f{\sqrt{\omega_{\rm out}}}{\sqrt{\omega_{\rm in}}}+\f{\sqrt{\omega_{\rm in}}}{\sqrt{\omega_{\rm out}}}\,\bigg],~~ 
\beta_\mathbf{k}=\f{1}{2}\bigg[\f{\sqrt{\omega_{\rm out}}}{\sqrt{\omega_{\rm in}}}-\f{\sqrt{\omega_{\rm in}}}{\sqrt{\omega_{\rm out}}}\,\bigg]\, 
\label{eq:bog3}. } 
For $w\neq 1/3$, we can approximate $\omega_{\rm in/out}=m_{\rm in/out}$ in the long-wavelength limit $\mathbf{k}^2\ll m_{\rm out}^2$. In that case the Bogoliubov coefficients are independent of momentum and we find the occupation number
\ee{n_\mathbf{k}\equiv\langle\text{in},0\vert\hat{n}_{\mathbf{k}}^{\rm out}\vert 0, \text{in}\rangle
=\vert \beta_\mathbf{k}\vert^2=\frac{(2-\sqrt{1-3w})^2}{8\sqrt{1-3w}}.
\label{eq:bogoBk}}
In the matter-dominated case ($w=0$) this reduces to $n_\mathbf{k}=1/8$.

In the radiation-dominated case ($w=1/3$), $\omega_{\rm out}=\vert\mathbf{k}\vert$. At long wavelengths, $\mathbf{k}^2\ll m_{\rm in}^2$, we find a momentum-dependent occupation number
\ee{n_\mathbf{k}=\vert \beta_\mathbf{k}\vert^2=\frac{\sqrt{3\xi} a_*H_*}{2\vert\mathbf{k}\vert},
\label{eq:bogoBkrad}}
where the star $*$ refers to values at the time of the transition and we have assumed $\xi\gg 1$. 

In reality, the change in the equation of state is not instantaneous, and the slower the transition, the fewer modes are excited. To quantify this, we make
use of Ref.~\cite{Bernard:1977pq} (see also Ref.~\cite{GarciaBellido:2001cb}), and model the time-dependence of the curvature mass term by
\ee{
m_{\rm curv}^2(\eta)=\frac{m^2_{\rm in}+m^2_{\rm out}}{2}
-\frac{m^2_{\rm in}-m^2_{\rm out}}{2}\tanh\nu\eta.
\label{eq:smoothm}}
Here $\nu$  controls the speed of the transition, with $\nu\rightarrow \infty$ and $\nu\rightarrow 0$ corresponding to infinitely fast and infinitely slow transitions, respectively. This case can also be solved analytically, and gives the occupation number
\ea{
n_\mathbf{k}=
\f{\sinh^2\left[\pi \left(\omega_{\rm out}-\omega_{\rm in}\right)/(2\nu)\right]}{\sinh(\pi\omega_{\rm in}/\nu)\sinh(\pi\omega_{\rm out}/\nu)}\, .\label{eq:occ}}
From the various limits of (\ref{eq:occ}) follow a number of physical implications: In the limit of large momentum, large mass or slow transition, i.e. $\mathbf{k}\rightarrow \infty$, $\xi\rightarrow\infty$ or $\nu\rightarrow 0$ the occupation number approaches zero exponentially. For a mode to receive excitations one must satisfy (assuming $m_{\rm in} > m_{\rm out}$)
\ee{\vert\mathbf{k}\vert\lesssim {\rm min}\left\{m_{\rm in}, \sqrt{(\nu/2\pi)^2 - m_{\rm out}^2}\,\right\}\, \label{eq:pc}.}

The excitation of the Higgs field due to the non-adiabatic behaviour at the end of inflation leads to a rapid growth of the fluctuations of the field.
To describe this  we calculate the variance of the field in the  "in" vacuum state, at time $\eta$ after the transition
\ea{
&\langle{\rm in},0\vert\hat{h}_{\rm out}^2\vert0,{\rm in}\rangle
\nonumber \\&=\int \f{d^3\vert \mathbf{k}\vert}{2\omega(2\pi)^3 a^2}\left[1+2\vert\beta_\mathbf{k}\vert^2
+2\alpha_\mathbf{k}\beta_\mathbf{k}\cos\Big(\int^\eta d\eta' \omega\Big)\right]\,.\label{eq:var0}}
These fluctuations are uncorrelated on superhorizon scales, and therefore they correspond to a stochastic background as opposed to the coherent Higgs condensate generated during inflation \cite{DeSimone:2012qr}.
This will lead to vacuum decay, 
if the field value averaged over a sufficiently large volume, which we conservatively consider to be the whole Hubble volume, exceeds the position of the potential barrier. Therefore we only include modes up to $\Lambda=aH$. We also neglect the vacuum part and coherently oscillating terms in (\ref{eq:var0}),
\ee{\langle \hat{h}^2\rangle_\Lambda\equiv\int_0^\Lambda \f{d\vert \mathbf{k}\vert\mathbf{k}^2\vert\beta_\mathbf{k}\vert^2}{2\pi^2 a^2\omega}\label{eq:var1}\, .}
Substituting Eqs.~(\ref{eq:bogoBk}) and (\ref{eq:bogoBkrad}), we get {for $\xi \gg 1$}
{\ee{\langle\hat{h}^2\rangle_{aH}
\simeq \left(\f{H}{2\pi}\right)^2 \!\!\times\begin{cases} (12\sqrt{3\xi})^{-1} \left(1-\f{m_{\rm in}}{m_{\rm out}}\right)^2,\quad &w\neq \f{1}{3}\\\sqrt{3\xi}\,a/a_*\,,&w=\f{1}{3}\,.
\end{cases}\label{eq:step}}}

From (\ref{eq:step}) and (\ref{eq:massratio}) we see that for $w\neq 1/3$ the superhorizon 
variance is suppressed by $\xi^{-1/2}$. This is however not the case for conformal equation of state $w=1/3$ due to inverse scaling $\xi^{1/2}$, whereby a large variance may be generated for $\xi \gg 1$ for a sufficiently rapid drop in $m_{\rm curv}$ \footnote{Taking in account backreaction due to increase of the effective mass (\ref{eff_mass}): $m_{\rm out}^2=m_{\rm eff}^2 = {6}\lambda \langle{h^2}\rangle$, where $\langle{h^2}\rangle \sim m_{\rm in}^2/(16 \pi^2 a^2)$ includes contributions from all modes up to $\Lambda \sim m_{\rm in}$, we find by using Eq.~(\ref{lambda_run}) for $\lambda$ a somewhat smaller estimate $\langle h^2\rangle_{aH} \simeq H^2/({18}\lambda_0\sqrt{3\xi})$ valid for $H < \Lambda_I$. In the region $H > \Lambda_I$ negative $\lambda$ would enhance the increase of the variance, however, in both regions the superhorizon variance is further constrained by the bounds (\ref{gravity_backreaction}) from gravity backreaction.}.

As an example, we assume a model where inflation ends within a time scale $\nu^{-1}\sim (400 H)^{-1}$ with an effective reheating equation of state $w=1/3$. Using (\ref{eq:step}) with $\xi \gtrsim 500$ we get for the magnitude of the generated variance: $\langle \hat{h}^2\rangle_{aH}{\gtrsim H^2}$. Therefore we may conclude that if inflation ends abruptly to a state with $w_{\rm out}\sim 1/3$, a variance larger than $\Lambda_I$ can be generated for sufficiently high inflationary scale $H \gtrsim \Lambda_I$, such that the fatal transition to the negative energy vacuum is likely triggered.

Above we assumed that the equation of state changes monotonically to its final value when inflation ends. However, this is not what happens in most models of inflation. Instead, the inflaton field oscillates coherently about its minimum, and this leads to a much stronger effect. For an inflationary model with a single inflaton field $\phi$ in potential $V(\phi)$, the scalar curvature $R$ is given by
\ee{R=\f{1}{M_{\rm pl}^2}\left[4 V(\phi)-\left(\frac{d\phi}{dt}\right)^2\right]\,.\label{eq:oscR}}
It is clear from this expression that $R$, and therefore also the curvature mass term $m_{\rm curv}^2$, oscillates between positive and negative values.
As a result, the Higgs field grows exponentially in a way that can be interpreted either as tachyonic or resonant growth.

Reheating via the curvature coupling $\xi$ was first studied in Ref.~\cite{Bassett:1997az} where it was named 'geometric reheating' (see also \cite{Tsujikawa:1999jh}).
{A similar tachyonic resonance effect present in reheating models with trilinear interactions was studied in \cite{Dufaux:2006ee}.}  
Assuming that the inflaton potential can be well approximated by the quadratic term \ee{V(\phi)=\f{m^2}{2}\phi^2\, ,\label{eq:infp}} we 
can write the solution as $\phi(t)=\Phi{\cos}( mt),$
where {$\Phi \approx \sqrt{6} H M_{\rm pl}/m$} is a slowly varying amplitude and $H\equiv2/(3t)$ as given by the equation of state for matter. Substituting this background solution into Eq.~(\ref{eq:oscR}) we obtain the Mathieu form of the mode equation (\ref{eq:mode}):
\ee{\f{d^2f(z)}{dz^2}+\bigg[A_k-2q\cos(2z)\bigg]f(z)=0,\qquad z=mt\,,\label{eq:mathieu}}
\ee{\nonumber A_k= \f{\mathbf{k}^2}{a^2m^2}+\xi \f{\Phi^2}{2 M_{\rm pl}^2},\qquad q=\f{3\Phi^2}{4 M_{\rm pl}^2}\bigg(\f{1}{4}-\xi\bigg)\, .}
Following the analysis of \cite{Dufaux:2006ee} we get for the occupation number after the first oscillation
\ee{n^{1}_\mathbf{k}=e^{2X_\mathbf{k}}\,,\qquad X_\mathbf{k}=\int_{\Delta z}\Omega_\mathbf{k}\,dz \approx \sqrt{\xi}\f{\Phi}{M_{\rm pl}}\, \label{eq:occapp},}
where $\Omega^2\equiv-\omega^2$ with $\omega^2$ being the term in the square brackets in (\ref{eq:mathieu}) and $\Delta z$ covers the time when $\omega^2<0$ during first oscillation. 
In the last form of (\ref{eq:occapp}) we have assumed that $\vert\mathbf{k}\vert \leq aH$ and that the amplitude $\Phi$ is roughly constant {($a\sim 1$}) during the oscillation. It is clear from (\ref{eq:occapp}) that for large $\xi$ the occupation number may become very large already during the first oscillation. Using (\ref{eq:occapp}) we get an estimate for the superhorizon variance after the first oscillation ($\vert \beta_{\mathbf{k}}\vert^2 = n^{1}_\mathbf{k}$)
\ee{\langle\hat{h}^2\rangle_{aH} \approx \int^{aH}_0 \f{d\vert\mathbf{k}\vert\, \mathbf{k}^2\vert \beta_{\mathbf{k}}\vert^2}{2\pi^2\omega_{\rm out}} 
\sim\left(\f{H}{2\pi}\right)^2\f{2\exp\left\{\sqrt{\xi}\f{2\Phi}{M_{\rm pl}}\right\}}{3\sqrt{3\xi}}\,,
\label{osc_variance}}
which is accurate for $\xi \geq 1/6+3/8\sim 0.5$. Below this region only superhorizon modes with $\mathbf{k}^2\leq k_{\rm cut}^2 \equiv a^2 H^2 6(\xi-3/8)$ receive tachyonic amplification. In this case, an accurate approximation for the superhorizon variance can be obtained by an expansion 
near the threshold point $\xi = 3/8$
\ee{\langle\hat{h}^2\rangle_{aH} \approx 16\sqrt{\f{3}{5}}\left(\f{H}{2\pi}\right)^2(\xi - 3/8)^{3/2}\,,\quad \f{3}{8} < \xi \lesssim \f{1}{2}\,.
\label{variance_thredshold}}
  
\begin{figure}
\includegraphics[width=0.48\textwidth]{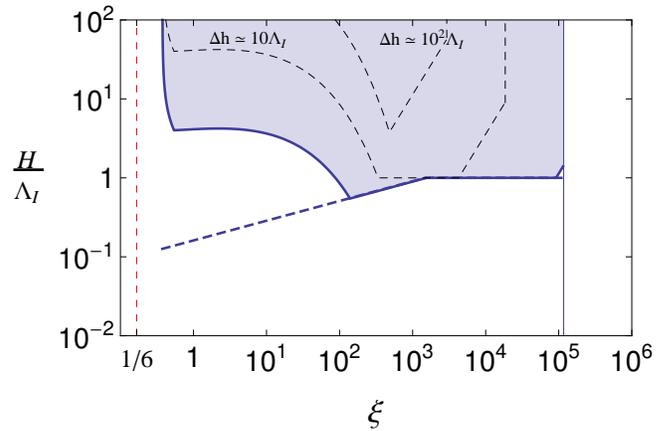}
\caption{\label{fig:regions}The estimated instability region (shaded, blue) where $\Delta h \gtrsim \Lambda_I$ after the first oscillation, given by Eqs.(\ref{osc_variance},\ref{variance_thredshold},\ref{self-int_backreaction},\ref{gravity_backreaction},\ref{instab_condition}). The dashed lines correspond to $\Delta h \gtrsim 10 \Lambda_I$ and $\Delta h \gtrsim 10^2\Lambda_I$. For the parameters we have used $\Phi = 0.3 M_{\rm pl}$ and $\Lambda_I = 10^{-7}M_{\rm pl}$.} 
\end{figure}

The exponential growth of the particle number and the variance is constrained by backreaction which eventually shuts off the tachyonic resonance and {makes the dynamics non-linear. We consider two backreaction effects: Higgs self-interaction and gravity. 

When considering the Higgs self-interaction, we adopt the convention that the RG-running parameters are evaluated at the scale $\mu = H$, which corresponds to the optimal choice in terms of RG-improved effective potential \cite{Herranen:2014cua} in case $H$ is the highest scale of the problem \footnote{More accurately, the renormalization scale should probably be chosen as $\mu = {\rm max}\{H,\langle\hat{h}\rangle,\sqrt{\langle\hat{h}^2\rangle}\}$, however, we have the checked that this choice (here $\langle\hat{h}\rangle=0$) would not alter the instability region of Figure \ref{fig:regions}.}. The RG-running self-coupling $\lambda(\mu)$ becomes negative at the scale of instability $\Lambda_I$ and therefore we may approximate ($\mu = H$):
\ee{\lambda(H) \simeq \lambda_0\, {\rm sign}(\Lambda_I - H)\,,\quad {\rm with} \quad \lambda_0 \approx 0.01\,.
\label{lambda_run}}
Hence, the backreaction from self-interactions is limited to the region $H < \Lambda_I$, while for $H > \Lambda_I$ the negative self-coupling enhances the tachyonic resonance and accelerates particle production. However, here we will not consider this enhancement but use the conservative estimate of (\ref{osc_variance}).

For $H<\Lambda_I$, the positive Higgs self-interaction generates an} effective mass
\ee{m_{\rm eff}^2 = m_{\rm curv}^2 + {6}\lambda_0 \langle\hat{h}^2\rangle\,.
\label{eff_mass}}
The resonance is shut off once (see also \cite{Bassett:1997az,Greene:1997ge})
%\ee{
$A_k + \delta A > 2|q|\,$
%,\label{resonance_shut}}
with $\delta A = {6}\lambda\langle\hat{h}^2\rangle / m^2$.
From this and Eqs.~(\ref{lambda_run}) we then obtain a condition for maximal variance in the region $H < \Lambda_I$: $\langle\hat{h}^2\rangle \lesssim \xi H^2/\lambda_0$, including contributions from all tachyonic modes $\vert\mathbf{k}\vert \leq k_{\rm cut}$. The corresponding superhorizon variance with $\vert\mathbf{k}\vert \leq aH$ is related to the total variance by $\langle\hat{h}^2\rangle_{aH} \sim \langle\hat{h}^2\rangle/(2(3\xi)^{3/2})$ and is hence constrained by
\ee{\langle\hat{h}^2\rangle_{aH} \lesssim \frac{H^2}{6 \lambda_0 \sqrt{3\xi}}\qquad{\rm for}\qquad H < \Lambda_I\,,
\label{self-int_backreaction}}
where we have used $R \approx 3 H^2$
 for $w \approx 0$ assumed to be valid after oscillations (or in average during oscillations).

For gravity backreaction we need to compare the energy density contained in produced Higgs particles to total energy density $\rho \sim 3 H^2 M_{\rm pl}^2$. In {Gaussian} (Hartree) approximation with $\langle\hat{h}^4\rangle = 3 \langle\hat{h}^2\rangle^2$ we get for the Higgs energy density an estimate
\ee{\rho_{\rm Higgs} \simeq 24 \xi H^2\langle\hat{h}^2\rangle + {6}\lambda\langle\hat{h}^2\rangle^2\,,\label{energy_dens}}
whereby we find the following constraints for the superhorizon variance for the Higgs energy density to remain sub-dominant during oscillations \footnote{We will not discuss in detail the cancellation effect when the two terms in (\ref{energy_dens}) have roughly equal magnitudes and $\lambda < 0$, however, we have checked that this effect alters the instability region in Figure \ref{fig:regions} only very mildly}:
\ee{\langle\hat{h}^2\rangle_{aH} \lesssim \frac{2 M_{\rm pl}^2}{11 (3\xi)^{5/2}}\quad{\rm and}\quad \langle\hat{h}^2\rangle_{aH} \lesssim \frac{H M_{\rm pl}}{2({2}\lambda_0)^{1/2}(3\xi)^{3/2}}\,.
\label{gravity_backreaction}}
 
We estimate that the probability for a potentially catastrophic transition to the negative energy vacuum is significant once the superhorizon variance has exceeded the instability scale $\Lambda_I \sim 10^{11}$ GeV of the Higgs effective potential:
\ee{\Delta h \equiv \sqrt{\langle\vert\hat{H}\vert^2\rangle_{aH}} = \sqrt{4\langle\hat{h}^2\rangle_{aH}} > \Lambda_I\,.
\label{instab_condition}}
We show the corresponding region after the first oscillation as blue in Figure \ref{fig:regions}. On the left, at $\xi\lesssim 10^2$, the increase of the variance is limited by the weakness of the resonance (\ref{osc_variance}), although the non-shaded area above the dashed blue line would be mostly filled by the subsequent 3-4 oscillations. Further to the left, the effect becomes even weaker and disappears completely at $\xi=3/8$. The bottom half of the plot is excluded by backreaction from the self-interactions. On the right, the backreaction from the gravity effects makes the linear approximation invalid, and more sophisticated methods are needed. Backreaction from the other SM degrees of freedom is not likely to change the qualitative picture as resonant amplification takes at least a few inflaton oscillations. A direct Higgs-inflaton coupling can enhance or weaken the effect depending on the model but would generally not eliminate it as long as the Higgs has a tachyonic phase during the first inflaton oscillation.

In conclusion, we have shown that for a sufficiently high inflationary scale $H \gtrsim \Lambda_I$, changing spacetime curvature can trigger a fatal transition to a negative energy vacuum, 
for a wide range of the curvature coupling $1 \lesssim \xi \lesssim 10^5$.
This conclusion applies to both rapid reheating and oscillating inflaton.
Combining this with the lower bound $\xi_{\rm EW}\gtrsim 0.1$ imposed by stability during inflation~\cite{Herranen:2014cua}, we find that the minimal scenario of Standard Model in a background of conventional high-scale inflation constrains the coupling $\xi$ to be close to its conformal value $\xi=1/6$.
 
\acknowledgments{MH is supported by the Villum Foundation Grant No. YIP/VKR022599, TM by the Osk. Huttunen foundation, SN by the Academy of Finland grant 257532, and AR by STFC grant ST/J0003533/1. The research leading to these results has received funding from the European Research Council under the European Union's Horizon 2020 program (ERC Grant Agreement no.648680).}


\begin{thebibliography}{99}
 
\bibitem{Degrassi:2012ry}

  G.~Degrassi, S.~Di Vita, J.~Elias-Miro, J.~R.~Espinosa, G.~F.~Giudice, G.~Isidori and A.~Strumia,
  %``Higgs mass and vacuum stability in the Standard Model at NNLO,''
  JHEP {\bf 1208} (2012) 098
  [arXiv:1205.6497 [hep-ph]];
  %%CITATION = ARXIV:1205.6497;%%
  %98 citations counted in INSPIRE as of 12 Jun 2013  
  
 %\cite{Bezrukov:2012sa}
\bibitem{Bezrukov:2012sa}
  F.~Bezrukov, M.~Y.~Kalmykov, B.~A.~Kniehl and M.~Shaposhnikov,
  %``Higgs Boson Mass and New Physics,''
  JHEP {\bf 1210} (2012) 140
  [arXiv:1205.2893 [hep-ph]];
  %%CITATION = ARXIV:1205.2893;%%
  %255 citations counted in INSPIRE as of 23 sept. 2015 
  
  %\cite{Buttazzo:2013uya}
  D.~Buttazzo, G.~Degrassi, P.~P.~Giardino, G.~F.~Giudice, F.~Sala, A.~Salvio and A.~Strumia;
  %``Investigating the near-criticality of the Higgs boson,''
  JHEP {\bf 1312} (2013) 089
  [arXiv:1307.3536].
  %%CITATION = ARXIV:1307.3536;%%
  %121 citations counted in INSPIRE as of 28 Jun 2014

\bibitem{Kobakhidze:2013tn}
%\cite{Espinosa:2007qp}
%\bibitem{Espinosa:2007qp}
  J.~R.~Espinosa, G.~F.~Giudice and A.~Riotto,
  %``Cosmological implications of the Higgs mass measurement,''
  JCAP {\bf 0805} (2008) 002
  [arXiv:0710.2484 [hep-ph]];

  O.~Lebedev,
  %``On Stability of the Electroweak Vacuum and the Higgs Portal,''
  Eur.\ Phys.\ J.\ C {\bf 72} (2012) 2058
  [arXiv:1203.0156 [hep-ph]];
  %%CITATION = ARXIV:1203.0156;%%
  %82 citations counted in INSPIRE as of 20 juil. 2015

  O.~Lebedev and A.~Westphal,
  %``Metastable Electroweak Vacuum: Implications for Inflation,''
  Phys.\ Lett.\ B {\bf 719} (2013) 415
  [arXiv:1210.6987 [hep-ph]];
  %%CITATION = ARXIV:1210.6987;%%
  %30 citations counted in INSPIRE as of 20 juil. 2015
  
  %%CITATION = ARXIV:0710.2484;%%
  %124 citations counted in INSPIRE as of 10 juin 2015
  A.~Kobakhidze and A.~Spencer-Smith,
  %``Electroweak Vacuum (In)Stability in an Inflationary Universe,''
  Phys.\ Lett.\ B {\bf 722} (2013) 130
  [arXiv:1301.2846 [hep-ph]];
  %%CITATION = ARXIV:1301.2846;%%
  %31 citations counted in INSPIRE as of 24 May 2015

%\cite{Fairbairn:2014zia}
  M.~Fairbairn and R.~Hogan,
  %``Electroweak Vacuum Stability in light of BICEP2,''
  Phys.\ Rev.\ Lett.\  {\bf 112} (2014) 201801
  [arXiv:1403.6786 [hep-ph]];
  %%CITATION = ARXIV:1403.6786;%%
  %27 citations counted in INSPIRE as of 24 May 2015

%\cite{Kobakhidze:2014xda}
  A.~Kobakhidze and A.~Spencer-Smith,
  %``The Higgs vacuum is unstable,''
  arXiv:1404.4709 [hep-ph];
  %%CITATION = ARXIV:1404.4709;%%
  %28 citations counted in INSPIRE as of 24 May 2015

%\cite{Enqvist:2014bua}
  K.~Enqvist, T.~Meriniemi and S.~Nurmi,
  %``Higgs Dynamics during Inflation,''
  JCAP {\bf 1407} (2014) 025
  [arXiv:1404.3699 [hep-ph]];
  %%CITATION = ARXIV:1404.3699;%%
  %21 citations counted in INSPIRE as of 24 May 2015


%\cite{Hook:2014uia}
  A.~Hook, J.~Kearney, B.~Shakya and K.~M.~Zurek,
  %``Probable or Improbable Universe? Correlating Electroweak Vacuum Instability with the Scale of Inflation,''
  JHEP {\bf 1501} (2015) 061
  [arXiv:1404.5953 [hep-ph]];
  %%CITATION = ARXIV:1404.5953;%%
  %20 citations counted in INSPIRE as of 24 May 2015

%\cite{Spencer-Smith:2014woa}
  A.~Spencer-Smith,
  %``Higgs Vacuum Stability in a Mass-Dependent Renormalisation Scheme,''
  arXiv:1405.1975 [hep-ph];
  %%CITATION = ARXIV:1405.1975;%%
  %23 citations counted in INSPIRE as of 04 juin 2015

  K.~Kamada,
  %``Inflationary cosmology and the standard model Higgs with a small Hubble induced mass,''
  Phys.\ Lett.\ B {\bf 742} (2015) 126
  [arXiv:1409.5078 [hep-ph]].
  %%CITATION = ARXIV:1409.5078;%%
  %7 citations counted in INSPIRE as of 24 May 2015

  F.~Bezrukov and M.~Shaposhnikov,
  %``Why should we care about the top quark Yukawa coupling?,''
  J.\ Exp.\ Theor.\ Phys.\  {\bf 120} (2015) 335
   [Zh.\ Eksp.\ Teor.\ Fiz.\  {\bf 147} (2015) 389]
  [arXiv:1411.1923 [hep-ph]];
  %%CITATION = ARXIV:1411.1923;%%
  %9 citations counted in INSPIRE as of 24 May 2015

  F.~Bezrukov, J.~Rubio and M.~Shaposhnikov,
  %``Living beyond the edge: Higgs inflation and vacuum metastability,''
  arXiv:1412.3811 [hep-ph];
  %%CITATION = ARXIV:1412.3811;%%
  %14 citations counted in INSPIRE as of 24 May 2015  

  S.~Di Chiara, V.~Keus and O.~Lebedev,
  %``Stabilizing the Higgs potential with a Z$'$,''
  Phys.\ Lett.\ B {\bf 744} (2015) 59
  [arXiv:1412.7036 [hep-ph]];
  %%CITATION = ARXIV:1412.7036;%%
  %1 citations counted in INSPIRE as of 04 Jun 2015

%\cite{Shkerin:2015exa}
  A.~Shkerin and S.~Sibiryakov,
  %``On stability of electroweak vacuum during inflation,''
  arXiv:1503.02586 [hep-ph];
  %%CITATION = ARXIV:1503.02586;%%
  %4 citations counted in INSPIRE as of 24 May 2015
 
  %\cite{Espinosa:2015qea}
  J.~R.~Espinosa, G.~F.~Giudice, E.~Morgante, A.~Riotto, L.~Senatore, A.~Strumia and N.~Tetradis,
  %``The cosmological Higgstory of the vacuum instability,''
  arXiv:1505.04825 [hep-ph].
  %%CITATION = ARXIV:1505.04825;%%

  %\cite{Herranen:2014cua}
\bibitem{Herranen:2014cua}
  M.~Herranen, T.~Markkanen, S.~Nurmi and A.~Rajantie,
  %``Spacetime curvature and the Higgs stability during inflation,''
  Phys.\ Rev.\ Lett.\  {\bf 113} (2014) 21,  211102
  [arXiv:1407.3141 [hep-ph]].
  %%CITATION = ARXIV:1407.3141;%%
  %21 citations counted in INSPIRE as of 24 May 2015

\bibitem{Kearney:2015vba}
%\cite{Kearney:2015vba}
  J.~Kearney, H.~Yoo and K.~M.~Zurek,
  %``Is a Higgs Vacuum Instability Fatal for High-Scale Inflation?,''
  arXiv:1503.05193 [hep-th];
  %%CITATION = ARXIV:1503.05193;%%
  %2 citations counted in INSPIRE as of 24 May 2015
 
%\cite{Markkanen:2014poa}
\bibitem{Markkanen:2014poa}
  T.~Markkanen,
  %``Curvature induced running of the cosmological constant,''
  Phys.\ Rev.\ D {\bf 91} (2015) 12,  124011
  [arXiv:1412.3991 [gr-qc]].
  %%CITATION = ARXIV:1412.3991;%%
  %6 citations counted in INSPIRE as of 04 Jun 2015 
  
%\cite{Atkins:2012yn}
\bibitem{Atkins:2012yn}
  M.~Atkins and X.~Calmet,
  %``Bounds on the Nonminimal Coupling of the Higgs Boson to Gravity,''
  Phys.\ Rev.\ Lett.\  {\bf 110} (2013) 051301
  [arXiv:1211.0281 [hep-ph]].
  %%CITATION = ARXIV:1211.0281;%%
  %15 citations counted in INSPIRE as of 14 Nov 2014

%\cite{Ade:2015tva}
\bibitem{Ade:2015tva}
  P.~A.~R.~Ade {\it et al.}  [BICEP2 and Planck Collaborations],
  %``Joint Analysis of BICEP2/$Keck  Array$ and $Planck$ Data,''
  Phys.\ Rev.\ Lett.\  {\bf 114} (2015) 10,  101301
  [arXiv:1502.00612 [astro-ph.CO]].
  %%CITATION = ARXIV:1502.00612;%%
  %107 citations counted in INSPIRE as of 24 May 2015

\bibitem{Bezrukov:2007ep}
  F.~L.~Bezrukov and M.~Shaposhnikov,
  %``The Standard Model Higgs boson as the inflaton,''
  Phys.\ Lett.\ B {\bf 659} (2008) 703
  [arXiv:0710.3755 [hep-th]].
  %%CITATION = ARXIV:0710.3755;%%
  %394 citations counted in INSPIRE as of 09 Apr 2014

\bibitem{DeSimone:2012qr}
  A.~De Simone and A.~Riotto,
  %``Cosmological Perturbations from the Standard Model Higgs,''
  JCAP {\bf 1302} (2013) 014
  [arXiv:1208.1344 [hep-ph]];
  %%CITATION = ARXIV:1208.1344;%%
  %14 citations counted in INSPIRE as of 10 juin 2015

%\bibitem{Kusenko}
  L.~Pearce, L.~Yang, A.~Kusenko and M.~Peloso,
  %``Leptogenesis Via Neutrino Production During Higgs Relaxation,''
  arXiv:1505.02461 [hep-ph];
  %%CITATION = ARXIV:1505.02461;%%

  K.~Enqvist, T.~Meriniemi and S.~Nurmi,
  %``Generation of the Higgs Condensate and Its Decay after Inflation,''
  JCAP {\bf 1310} (2013) 057
  [arXiv:1306.4511 [hep-ph]];
  %%CITATION = ARXIV:1306.4511;%%
  %15 citations counted in INSPIRE as of 24 May 2015

  K.~Enqvist, S.~Nurmi and S.~Rusak,
  %``Non-Abelian dynamics in the resonant decay of the Higgs after inflation,''
  JCAP {\bf 1410} (2014) 10,  064
  [arXiv:1404.3631 [astro-ph.CO]];
  %%CITATION = ARXIV:1404.3631;%%
  %7 citations counted in INSPIRE as of 24 May 2015  

  D.~G.~Figueroa, J.~Garcia-Bellido and F.~Torrenti,
  %``The Decay of the Standard Model Higgs after Inflation,''
  arXiv:1504.04600 [astro-ph.CO].
  %%CITATION = ARXIV:1504.04600;%%
  %2 citations counted in INSPIRE as of 24 May 2015

%\cite{Ford:1986sy}
\bibitem{Ford:1986sy}
  L.~H.~Ford,
  %``Gravitational Particle Creation and Inflation,''
  Phys.\ Rev.\ D {\bf 35} (1987) 2955.
  %%CITATION = PHRVA,D35,2955;%%
  %206 citations counted in INSPIRE as of 02 Nov 2015


%\cite{Bassett:1997az}
\bibitem{Bassett:1997az}
  B.~A.~Bassett and S.~Liberati,
  %``Geometric reheating after inflation,''
  Phys.\ Rev.\ D {\bf 58} (1998) 021302
   [Phys.\ Rev.\ D {\bf 60} (1999) 049902]
  [hep-ph/9709417].
  %%CITATION = HEP-PH/9709417;%%
  %43 citations counted in INSPIRE as of 21 May 2015

%\cite{Tsujikawa:1999jh}
\bibitem{Tsujikawa:1999jh}
  S.~Tsujikawa, K.~i.~Maeda and T.~Torii,
  %``Resonant particle production with nonminimally coupled scalar fields in preheating after inflation,''
  Phys.\ Rev.\ D {\bf 60} (1999) 063515
  [hep-ph/9901306].
  %%CITATION = HEP-PH/9901306;%%
  %38 citations counted in INSPIRE as of 21 May 2015
  
%\cite{Parker:1968mv}
\bibitem{Parker:1968mv}
  L.~Parker,
  %``Particle creation in expanding universes,''
  Phys.\ Rev.\ Lett.\  {\bf 21} (1968) 562;
  %%CITATION = PRLTA,21,562;%%
  %335 citations counted in INSPIRE as of 21 May 2015

  L.~Parker,
  %``Quantized fields and particle creation in expanding universes. 1.,''
  Phys.\ Rev.\  {\bf 183} (1969) 1057;
  %%CITATION = PHRVA,183,1057;%%
  %669 citations counted in INSPIRE as of 21 May 2015


%\cite{Bernard:1977pq}
\bibitem{Bernard:1977pq}
  C.~W.~Bernard and A.~Duncan,
  %``Regularization and Renormalization of Quantum Field Theory in Curved Space-Time,''
  Annals Phys.\  {\bf 107} (1977) 201.
  %%CITATION = APNYA,107,201;%%
  %59 citations counted in INSPIRE as of 21 May 2015


%\cite{GarciaBellido:2001cb}
\bibitem{GarciaBellido:2001cb}
  J.~Garcia-Bellido and E.~Ruiz Morales,
  %``Particle production from symmetry breaking after inflation,''
  Phys.\ Lett.\ B {\bf 536} (2002) 193
  [hep-ph/0109230].
  %%CITATION = HEP-PH/0109230;%%
  %61 citations counted in INSPIRE as of 21 May 2015

%\cite{Dufaux:2006ee}
\bibitem{Dufaux:2006ee}
  J.~F.~Dufaux, G.~N.~Felder, L.~Kofman, M.~Peloso and D.~Podolsky,
  %``Preheating with trilinear interactions: Tachyonic resonance,''
  JCAP {\bf 0607} (2006) 006
  [hep-ph/0602144].
  %%CITATION = HEP-PH/0602144;%%
  %60 citations counted in INSPIRE as of 21 May 2015

%\cite{Greene:1997ge}
\bibitem{Greene:1997ge} 
  B.~R.~Greene, T.~Prokopec and T.~G.~Roos,
  %``Inflaton decay and heavy particle production with negative coupling,''
  Phys.\ Rev.\ D {\bf 56}, 6484 (1997)
  [hep-ph/9705357].
  %%CITATION = HEP-PH/9705357;%%
  %84 citations counted in INSPIRE as of 25 May 2015
  
%\cite{Bednyakov:2015sca}
\bibitem{Bednyakov:2015sca}
  A.~V.~Bednyakov, B.~A.~Kniehl, A.~F.~Pikelner and O.~L.~Veretin,
  %``Stability of the Electroweak Vacuum: Gauge Independence and Advanced Precision,''
  arXiv:1507.08833 [hep-ph]:
  %%CITATION = ARXIV:1507.08833;%%
  %5 citations counted in INSPIRE as of 23 sept. 2015
  
  S.~Frixione and A.~Mitov,
  %``Determination of the top quark mass from leptonic observables,''
  JHEP {\bf 1409}, 012 (2014)
  [arXiv:1407.2763 [hep-ph]].
  %%CITATION = ARXIV:1407.2763;%%
  %6 citations counted in INSPIRE as of 21 sept. 2015  
  
 %\cite{Jackiw:1974cv}
\bibitem{Jackiw:1974cv}
  R.~Jackiw,
  %``Functional evaluation of the effective potential,''
  Phys.\ Rev.\ D {\bf 9} (1974) 1686;
  %%CITATION = PHRVA,D9,1686;%%
  %801 citations counted in INSPIRE as of 31 May 2015
  
  L.~Di Luzio and L.~Mihaila,
  %``On the gauge dependence of the Standard Model vacuum instability scale,''
  JHEP {\bf 1406} (2014) 079
  [arXiv:1404.7450 [hep-ph]];
  %%CITATION = ARXIV:1404.7450;%%
  %20 citations counted in INSPIRE as of 31 May 2015  


  A.~Andreassen, W.~Frost and M.~D.~Schwartz;
  %``Consistent Use of the Standard Model Effective Potential,''
  Phys.\ Rev.\ Lett.\  {\bf 113} (2014) 24,  241801
  [arXiv:1408.0292 [hep-ph]].
  %%CITATION = ARXIV:1408.0292;%%
  %16 citations counted in INSPIRE as of 04 juin 2015
  
   A.~Andreassen, W.~Frost and M.~D.~Schwartz,
  %``Consistent Use of Effective Potentials,''
  Phys.\ Rev.\ D {\bf 91} (2015) 1,  016009
  [arXiv:1408.0287 [hep-ph]].
  %%CITATION = ARXIV:1408.0287;%%
  %8 citations counted in INSPIRE as of 31 May 2015  


\end{thebibliography}
\end{document}